\documentclass[11pt]{article}
\usepackage{amsmath,amssymb,color,epsfig,cite}
\usepackage{graphicx}
\usepackage{setspace}

\usepackage{mathrsfs}

\usepackage{amsfonts}
\newcommand{\hoch}[1]{$\, ^{#1}$}

\makeatletter
\@addtoreset{equation}{section}
\makeatother

\newcommand{\be}{\begin{equation}}
\newcommand{\ee}{\end{equation}}
\newcommand{\bea}{\setlength\arraycolsep{2pt} \begin{eqnarray}}
\newcommand{\eea}{\end{eqnarray}}

\newcommand{\half}{{\textstyle{\frac{1}{2}}}}

\def\ndelta{\delta\hspace{-0.50em}\slash\hspace{-0.05em} }

\def\ft#1#2{{\textstyle{\frac{\scriptstyle #1}{\scriptstyle #2} } }}
\def\fft#1#2{{\frac{#1}{#2}}}

\def\0{{\sst{(0)}}}
\def\1{{\sst{(1)}}}
\def\2{{\sst{(2)}}}
\def\3{{\sst{(3)}}}
\def\4{{\sst{(4)}}}
\def\5{{\sst{(5)}}}
\def\6{{\sst{(6)}}}
\def\7{{\sst{(7)}}}
\def\8{{\sst{(8)}}}
\def\sst#1{{\scriptscriptstyle #1}}

\def\del{{\partial}}

\def\scri{\mathscr{I}}

\begin{document}

\begin{flushright}
\hfill{\hfill{MI-TH-1812}}

\end{flushright}

\vspace{15pt}
\begin{center}
{\Large {\bf New dual gravitational 
charges}}

\vspace{15pt}
{\bf Hadi Godazgar\hoch{1}, Mahdi Godazgar\hoch{2} and 
C.N. Pope\hoch{3,4}}

\vspace{10pt}

\hoch{1} {\it Max-Planck-Institut f\"ur Gravitationsphysik (Albert-Einstein-Institut), \\
M\"ühlenberg 1, D-14476 Potsdam, Germany.}

\vspace{10pt}

\hoch{2} {\it Institut f\"ur Theoretische Physik,\\
Eidgen\"ossische Technische Hochschule Z\"urich, \\
Wolfgang-Pauli-Strasse 27, 8093 Z\"urich, Switzerland.}

\vspace{10pt}

\hoch{3} {\it George P. \& Cynthia Woods Mitchell  Institute
for Fundamental Physics and Astronomy,\\
Texas A\&M University, College Station, TX 77843, USA.}

\vspace{10pt}

\hoch{4}{\it DAMTP, Centre for Mathematical Sciences,\\
 Cambridge University, Wilberforce Road, Cambridge CB3 OWA, UK.}

 \vspace{15pt}
\today

\vspace{20pt}

\underline{ABSTRACT}
\end{center}

\noindent We show that there are a further infinite number of, previously unknown, supertranslation charges. These can be viewed as duals of the known BMS charges corresponding to supertranslations. In Newman-Penrose language, these new supertranslation charges roughly correspond to the imaginary part of the leading term in $\psi_2$. We find these charges by dualising the Barnich-Brandt asymptotic charges and argue that this prescription gives rise to new \emph{bona fide} charges at null infinity.

\thispagestyle{empty}

\vfill
E-mails: hadi.godazgar@aei.mpg.de, godazgar@phys.ethz.ch, pope@physics.tamu.edu

\pagebreak

\section{Introduction}

Recently, the relation between BMS symmetry and Newman-Penrose charges at null infinity of asymptotically flat spacetime has been made explicit in linear and non-linear gravity~\cite{conde, fakenews}, as well as electromagnetism~\cite{Campiglia:2018dyi}. While BMS charges are strictly defined at null infinity, and in particular include the Bondi 4-momentum, it has been shown that other charges can be defined by extending the definition of BMS charges into the bulk and it is these extended BMS charges that encompass some of the Newman-Penrose charges. In linearised gravity, at each order in a $1/r$ expansion away from null infinity the Newman-Penrose charges are components of the Weyl scalar $\psi_0$ in a $1/r$ expansion~\cite{NP}---the real parts of which correspondingly extend the notion of BMS charges as a  $1/r$ expansion into the bulk~\cite{conde}. Furthermore, the same picture holds in the non-linear theory, where an extension of the BMS charges using the Barnich-Brandt prescription \cite{BB} as a $1/r$ expansion away from null infinity is shown to include five of the ten non-linear Newman-Penrose charges~\cite{fakenews}. It remains an open question whether the extension of the BMS charges into the bulk can be further enlarged such that they contain the imaginary parts of the Newman-Penrose charges. In this paper we will not resolve this question in the general setting of extended BMS charges but show that already at the level of the standard BMS charges something has been hitherto missed. 

At leading order, the BMS charges can be derived from the Barnich-Brandt 
formalism \cite{BarTro}. By making a particular choice of the 
supertranslation parameter $s(\theta,\phi)$, namely choosing 
$l=0,1$ spherical harmonics,~\footnote{The supertranslation parameter 
describing a
diffeomorphism of a physical metric should, of course, be real.  It is
convenient to decompose a general such parameter $s(\theta,\varphi)$ as a
sum over spherical harmonics, which we may think of as the complete set
of (real) solutions of $\square\, s = -\ell(\ell+1)\, s$ on the unit sphere,
where $\ell=0,1,2,\cdots$.  It will always be understood that we are
taking $s(\theta,\phi)$ to be real.  Of course in practice it is often
convenient to work with the complex basis of spherical harmonics
$Y_{\ell m}(\theta,\phi)$. Whenever, in this paper, we speak of taking
$s(\theta,\phi)$ to be a harmonic $Y_{\ell m}(\theta,\phi)$, it should be
understood that really, we mean that $s$ is a real function
constructed as an appropriate linear combination of the complex
$Y_{\ell m}(\theta,\phi)$ harmonics.} the BMS charge can be shown to include 
the \emph{real} part~\footnote{To be precise, the real part of
$-1/(4 G)\int d\Omega \, s\, 
(\psi_2^0 + \sigma^0 \partial_u \bar{\sigma}^0)$,
where $s$ is any of the four linearly-independent real harmonics 
proportional to $Y_{0,0}$,
$Y_{1,0}$, $(Y_{1,1}-Y_{1,-1})$ or $i\,(Y_{1,1}+Y_{1,-1})$.} 
of the Bondi 4-momentum 
\cite{BarTro}
\begin{equation} \label{4mtm}
 P_{\ell,m} = - \frac{1}{2\sqrt{\pi}\, G}  
\int d\Omega\ Y_{\ell m}\; (\psi_2^0 + \sigma^0 \partial_u \bar{\sigma}^0),
\end{equation}
where $\ell= 0$ or 1, 
and $\psi_2^0$ and $\sigma^0$ are the leading terms in a $1/r$-expansion of the Weyl scalar $\psi_2$ and the shear $\sigma$, respectively:
\begin{equation}
	\psi_2^0 = \lim_{r\rightarrow\infty} r^3 \, \psi_{2}, \qquad \textrm{and} \qquad \sigma^{0} = \lim_{r\rightarrow\infty} r^2 \, \sigma.\label{psisigma}
\end{equation}
For $\ell=0$ or 1, the fact that the Barnich-Brandt prescription gives only the real part is not so troubling, since one can show that
\begin{equation} \label{impsi20}
 \Im(\psi_2^0 + \sigma^0 \partial_u \bar{\sigma}^0) = 
\Im(\bar{\eth}^2 \sigma^0).
\end{equation}
Now, $\bar{\eth}^2 Y_{\ell m} =0 $ for $\ell=0$ or 1, and so the 
imaginary part is a total derivative, which vanishes under the integral 
over the sphere.  

   If we consider instead an arbitrary 
supertranslation parameter, then 
\begin{equation} 
s(\theta, \phi)\, \Im(\psi_2^0 + \sigma^0 \partial_u \bar{\sigma}^0) = s(\theta, \phi)\, \Im(\bar{\eth}^2 \sigma^0)
\end{equation}
is no longer a total derivative when $\ell\ge2$. Thus, one may ask if there is 
a sense in which the Barnich-Brandt prescription is only giving half of the 
asymptotic charges when $\ell\ge2$ (i.e.~only the real part of the complex
generalised charge $-1/(4\pi G)\int d\Omega \, s\, 
(\psi_2^0 + \sigma^0 \partial_u \bar{\sigma}^0)$).  
It is this question that we shall 
address in this paper.  Indeed, as we shall show, we may define 
an infinite number of extra supertranslation charges. These charges are 
obtained by considering the ``dual'' of the Barnich-Brandt asymptotic charge, 
which is the analogue of considering the field strength and its dual in the 
case of electromagnetism~\cite{Campiglia:2018dyi}.  In a 
gravitational context, it is analogous to getting a NUT charge by 
dualising the Bondi mass~\cite{Ramaswamy} or Komar mass~\cite{Bossard:2008sw}.

In section \ref{sec:EM}, we consider for illustrative purposes the simpler case of electromagnetism and show how the usual electric and magnetic charges can be viewed as the real and imaginary parts of the Newman-Penrose charge, respectively.  We extend this analogy to the gravitational case in section \ref{sec:gravity} and find that one can define dual gravitational charges corresponding to the supertranslation generators of the BMS group at null infinity.  We conclude 
with some comments in section \ref{sec:dis}.

\section{Electromagnetism} \label{sec:EM}

We begin by considering the simpler case of electromagnetism on flat Minkowski spacetime \cite{Campiglia:2018dyi}, with metric given in outgoing Eddington-Finkelstein coordinates $(u,r,x^I=\{\theta,\phi\})$ by
\begin{equation} \label{Mink}
 d s^2 = - du^2 - 2 du dr + r^2 \omega_{IJ} \, dx^I dx^J.
\end{equation}
A convenient choice of complex null frame $e_\mu{}^a=(\ell^a,n^a,m^a,\bar{m}^a)$ is given by
\begin{align}
 &\ell = \frac{\partial}{\partial r}, \hspace{10.9mm} n =  \frac{\partial}{\partial u} - \frac{1}{2} \frac{\partial}{\partial r} , \hspace{15mm} m = \frac{\hat{m}^I}{r}  \frac{\partial}{\partial x^I}, \notag\\
 &\ell^\flat = - du, \qquad n^{\flat} = - \Big( dr + \frac{1}{2}  du \Big), \qquad m^{\flat} = r\, \hat{m}_I\, dx^I, \notag \\
 & \hat{m} = \frac{1}{\sqrt{2}} \left(\frac{\partial}{\partial \theta} + \frac{i}{\sin \theta} \frac{\partial}{\partial \phi} \right), \qquad \hat{m}^\flat = \frac{1}{\sqrt{2}} \left(d \theta + i \sin \theta d \phi \right).
 \label{Mink:frame}
\end{align}

Following Barnich and Brandt \cite{BB}, we define the \emph{electric} charge to be\footnote{Note that in the case of electromagnetism, the Barnich-Brandt charge is integrable.  This is not the case in non-linear gravity due to Bondi news (or more generally fake news \cite{fakenews}) at null infinity.}
\begin{equation} \label{EMcharge}
 \mathcal{Q}_c = \frac{1}{4\pi} \int_S c \star F = \frac{1}{4\pi} \int_S d \Omega\ c\, r^2\,  F_{01},
\end{equation}
where $c(x)$ is an arbitrary function on the 2-sphere corresponding to the asymptotic symmetry for electromagnetism and we use the notation that for some arbitrary covector $V$
\begin{equation}
 \ell^a V_a \equiv V_0 = - V^1,\qquad n^a V_a \equiv V_1 = -V^0,\qquad m^a V_a \equiv V_m=V^{\bar{m}}.
\end{equation}

Contrast the above expression with the Newman-Penrose charge \cite{NP}, generalised to include a constant $c$
\begin{equation} \label{EM:NP}
  \mathcal{Q}^{(NP)}_c = \lim_{r \rightarrow \infty } \frac{1}{2 \pi} \int_S \ c\, r^2 \ \Phi_1,
\end{equation}
where 
\begin{equation}
 \Phi_1 = \frac{1}{2} (F_{01} + F_{m \bar{m}})
\end{equation}
is a Newman-Penrose scalar corresponding to a particular component of the Maxwell field strength in the complex null frame. We only take the leading Newman-Penrose charge and do not, here, consider a $1/r$-expansion in which case one could define a charge at every order.
We stress that what appears in integral \eqref{EM:NP} is the \emph{complex} Newman-Penrose scalar $\Phi_1$ multiplied by a constant.  Note that the real part of $\Phi_1$ is given by $F_{01}$, which corresponds to the expression in the Barnich-Brandt integral \eqref{EMcharge}.  What about the imaginary part of the generalised Newman-Penrose charge given by $F_{m\bar{m}}$?

As emphasised above, the Barnich-Brandt integral with $c=1$ corresponds to the electric charge.  Correspondingly, the asymptotic \emph{magnetic} charge may be defined as

\begin{equation}
  \tilde{\mathcal{Q}}_c = \frac{1}{4\pi} \int_S c\, F = \frac{1}{4\pi} \int_S d \Omega\ i\, c\, r^2\,  F_{m \bar{m}}.
\end{equation}
Given this we conclude that
\begin{equation}
 \mathcal{Q}^{(NP)}_c = \mathcal{Q}_c - i \tilde{\mathcal{Q}}_c,
\end{equation}
i.e.\ the generalised Newman-Penrose charge contains information about both the electric \emph{and} magnetic charge.

\paragraph{Aside}  It may be argued that for $c=1$, $\tilde{\mathcal{Q}} = 0$, as follows: Stokes' theorem implies that
\begin{equation}
 \tilde{\mathcal{Q}} = \frac{1}{4\pi} \int_S \, F =  \frac{1}{4\pi} \int_\Sigma \, d F = 0
\end{equation}
by the Bianchi identity.  However, this result follows if null infinity is the only boundary of the spacetime.  On a black hole background this result need not hold as the magnetic charge at infinity is equal and opposite to a contribution to the integral from the horizon. 

\section{Gravity} \label{sec:gravity}

As is to be expected, the case of gravity is more intricate compared to the electromagnetic case.  Starting from an asymptotically flat spacetime \cite{bondi, sachs}, which we define to be a spacetime for which there exist Bondi coordinates $(u,r,x^I=\{\theta,\phi\})$ in which the metric takes the form 
\begin{equation} \label{AF}
 d s^2 = - F e^{2 \beta} du^2 - 2 e^{2 \beta} du dr + 
r^2 h_{IJ} \, (dx^I - C^I du) (dx^J - C^J du)
\end{equation}
with the metric functions satisfying the following fall-off conditions at large $r$
\begin{align}
 F(u,r,x^I) &= 1 + \frac{F_0(u,x^I)}{r} +  o(r^{-1}), \notag \\[2mm]
 \beta(u,r,x^I) &= \frac{\beta_0(u,x^I)}{r^2} + o(r^{-3}), \notag \\[2mm] 
 C^I(u,r,x^I) &= \frac{C_0^I(u,x^I)}{r^2} + o(r^{-2}), \notag \\[2mm] \label{met:falloff}
	h_{IJ}(u,r,x^I) &= \omega_{IJ} + \frac{C_{IJ}(u,x^I)}{r} + o(r^{-1}),
\end{align}
where $\omega_{IJ}$ is the standard metric on the round 2-sphere with coordinates $x^I=\{\theta, \phi\}$.  Moreover, residual gauge freedom allows us to require that
\begin{equation} \label{det:h}
 h =\omega,
\end{equation}
where $h \equiv \textup{det}(h_{IJ})$ and $\omega \equiv \textup{det}(\omega_{IJ}) =\sin^2\theta$.  Furthermore, we assume that 
\begin{equation} \label{falloff:matter}
 T_{0m} = o(r^{-3})
\end{equation}
so that the Einstein equation then implies that \cite{BarTro,fakenews}
\begin{equation} \label{C0eqn}
C_0^I = -\half D_J C^{IJ},
\end{equation}
where $D_I$ is the covariant derivative compatible with the metric on the round 2-sphere $\omega_{IJ}.$

The BMS charge is defined as~\cite{BB, BarTro} 
\begin{equation} \label{BB:charge}
  \ndelta \mathcal{Q} = \frac{1}{8 \pi G} \lim_{r\rightarrow\infty} \int_{S}\,\star H = \frac{1}{8 \pi G} \lim_{r\rightarrow\infty}  \int_{S} d \Omega\ r^2 e^{2\beta} H^{ur} ,
\end{equation}
where 
\begin{equation} \label{H}
 H = \frac{1}{2} \Big\{ \xi_b g^{cd} \nabla_a \delta g_{cd} -\xi_b \nabla^c \delta g_{ac} +\xi^c \nabla_b \delta g_{ac} + \frac{1}{2} g^{cd} \delta g_{cd} \nabla_b \xi_a + \frac{1}{2} \delta g_{bc} (\nabla_a \xi^c - \nabla^c \xi_a) \Big\} dx^a \wedge dx^b
\end{equation}
and the notation $\ndelta$ is used to signify the fact that the expression is not necessarily integrable.  The asymptotic symmetry generator
\begin{equation} \label{BMSgen}
 \xi = s\, \partial_u +   \int dr \frac{e^{2\beta}}{r^2} h^{IJ} D_{J} s \  \partial_I - \frac{r}{2} \left( D_I \xi^I - C^I D
 _I s \right) \partial_r
\end{equation}
with $s(x)$ an arbitrary function on the 2-sphere.

Given the boundary conditions \eqref{met:falloff}, the BMS charge  \eqref{BB:charge} reduces to \cite{BarTro}
\begin{equation} \label{I0}
 \ndelta \mathcal{Q} = \frac{1}{16 \pi G} \int_{S} d \Omega\  \Bigg[ \delta \big( -2 s \, F_{0} \big) + \frac{s}{2} \partial_u C_{IJ} \delta C^{IJ} \Bigg].
\end{equation}
The integrable part of the charge is given by
\begin{equation} \label{BB:Q0}
 \mathcal{Q}^{(int)} =-  \frac{1}{8 \pi G} \int_{S} d \Omega\ s \, F_{0},
\end{equation}
while the non-integrable part can be interpreted as the existence of Bondi flux at null infinity, which prevents the conservation of the charge along null infinity.

Alternatively, we may define the 
charge 
\begin{equation} \label{BMS:charge}
 \mathcal{Q} = - \frac{1}{4\pi G} \int d\Omega\ s\; 
	(\psi_2^0 + \sigma^0 \partial_u \bar{\sigma}^0),
\end{equation}
where $\psi_2^0$ and $\sigma^0$ are defined in (\ref{psisigma}).  As discussed
in \cite{NP} (see equation (4.8) or (4.17) of Ref.\ \cite{NP}), one has
\be
\del_u \, \mathcal{Q} = - \frac{1}{4\pi G} \int d\Omega\ s\; 
        \Big( |\del_u\, \sigma^0|^2  -\eth^2(\del_u\bar\sigma^0)\Big).
\label{Qdot}
\ee
Newman and Penrose only considered the case where $s$ is taken to be an $\ell=0$
or $\ell=1$ spherical harmonic $Y_{\ell m}$, since after integration by
parts on the second term one has a factor $\bar\eth^2\, Y_{\ell m}$, 
which vanishes identically. These $\ell=0$ and $\ell=1$ charges give the
Bondi-Sachs mass and 3-momentum respectively \cite{NP}.  In particular,
the $\ell=0$ Bondi mass (or more precisely energy) is seen to be a strictly non-increasing function of
$u$, which is conserved if and only if $\del_u\, \sigma^0=0$.  In
terms of the metric components defined in the expansions (\ref{met:falloff}),
one has 
\be
|\del_u\, \sigma^0|^2 = \ft18 N^{IJ}\, N_{IJ}\,,
\ee
where $N_{IJ}= \del_u\, C_{IJ}$ is the Bondi news tensor.  Thus
the Bondi-Sachs mass and 3-momentum are conserved if and only if the 
Bondi news tensor vanishes, signifying the absence of  
gravitational radiation at future null infinity $\scri^+$. 

More generally, we may allow the function $s$ in the charge 
(\ref{BMS:charge}) to be any arbitrary spherical harmonic, without the
restriction to $\ell=0$ or $\ell=1$, and we again have charges that are
conserved whenever the Bondi news tensor vanishes.\footnote{What one
loses, by considering the infinity of charges corresponding to $\ell\ge2$,
is that now the non-conservation when $N_{IJ}\ne0$ is no longer of a 
definite sign, since both the
$\eth^2(\del_u\, \sigma^0)$ and the $|\del_u\, \sigma^0|^2$ terms contribute
when $N_{IJ}\ne 0$.  See, however, appendix \ref{app:C}.}  Our focus in the remainder of this section will be
on showing how these more general charges \eqref{BMS:charge} are related to
Barnich-Brandt BMS charges, and a generalisation thereof.

Calculating $\psi_2^0$ and
$\sigma^0$ in terms of the metric expansion coefficients in (\ref{met:falloff}),
one finds 
\be
\psi_2^0 + \sigma^0 \partial_u \bar{\sigma}^0= \ft12 F_0 -
 \fft{i}{4}  D_I D_J\, \tilde C^{IJ},\label{NPintegrand}
\ee
and so the two expressions \eqref{BB:Q0} and \eqref{BMS:charge} are related by
\begin{equation}
 \mathcal{Q}^{(int)} = \Re \left( \mathcal{Q} \right),\label{BBrealNP}
\end{equation}
where we take $s$ to be an arbitrary function of $x^I$ in the definition 
of ${Q}$.
This is analogous to what we found before in section \ref{sec:EM}, namely, 
for the asymptotic symmetry chosen to give a global charge, the BMS 
charge is the real part of the more general charge that we have defined in equation \eqref{BMS:charge}.  

Noting that (\ref{BBrealNP}) has only provided a relation between the 
{\it real} part of the
charge (\ref{BMS:charge}) and the Barnich-Brandt charge (\ref{BB:Q0}),
and inspired by the electromagnetic example in the previous section, 
we are now led to 
consider the \emph{dual} or \emph{magnetic} Barnich-Brandt charge
\begin{equation} \label{BB:dual}
  \ndelta \tilde{\mathcal{Q}} =\frac{1}{8 \pi G} \lim_{r\rightarrow\infty}  \int_{S}\, H = \frac{1}{8 \pi G} \lim_{r\rightarrow\infty} \int_{S} d \Omega \ \frac{H_{\theta \phi}}{\sin \theta}
\end{equation}
with $H$ defined in equation \eqref{H}.  It remains to show that this defines a charge, namely that the quantity 
defined above vanishes on-shell. We show that this is the case in 
appendix \ref{app:charge}.

It is straightforward to show that (see appendix \ref{app:dualH})
\begin{equation} \label{dualH}
 \ndelta \tilde{\mathcal{Q}} = \frac{1}{16 \pi G} \int_{S} d \Omega \ \Bigg[ \delta \big( -  s D_I D_J \tilde{C}^{IJ} \big) + \frac{s}{2} \partial_u C_{IJ} \delta \tilde{C}^{IJ} \Bigg],
\end{equation}
where~\footnote{In fact, $C_K{}^{[I} \epsilon^{J]K} = 0,$ which can simply be shown using Schouten identities in two dimensions and the trace-free property of $C_{IJ}$.  Thus, $\tilde{C}^{IJ} = C_K{}^{I} \epsilon^{JK}$.}
\begin{equation} \label{Ctwist}
\tilde{C}^{IJ} = C_K{}^{(I} \epsilon^{J)K}, \qquad \epsilon_{IJ} =  
\begin{pmatrix}                                                                                0 & 1 \\ -1 & 0                                                              \end{pmatrix} \sin \theta.
\end{equation}
Note that the non-integrable term is closely analogous to that 
for $\ndelta \mathcal{Q}$, see equation \eqref{I0}, and it also
vanishes if the Bondi news vanishes.  The integrable part gives
rise to new charges
\begin{equation}  \label{BB:dQ0}
 \tilde{\mathcal{Q}}^{(int)} =- \frac{1}{16 \pi G} \int_{S} d \Omega\  s 
\, D_I D_J \tilde{C}^{IJ}
\end{equation}
that are conserved in the absence of Bondi news.  As can be seen from
(\ref{NPintegrand}),
\begin{equation}
  D_I D_J \tilde{C}^{IJ} = - 4\, 
\Im (\psi_2^0 + \sigma^0 \partial_u \bar{\sigma}^0),
\end{equation}
and so we have
\begin{equation}
 \mathcal{Q} = \mathcal{Q}^{(int)} - i \tilde{\mathcal{Q}}^{(int)}.
\end{equation}

Integrating by parts, $\tilde{\mathcal{Q}}^{(int)}$ in (\ref{BB:dQ0}) can
be rewritten as
\begin{equation}  \label{BB:dQ02}
 \tilde{\mathcal{Q}}^{(int)} =- \frac{1}{16 \pi G} \int_{S} d \Omega\ \, 
	(D_I D_J \,s) \, \tilde{C}^{IJ}.
\end{equation}
If $s$ is an $\ell=0$ or $\ell=1$ spherical harmonic, in which case
$s$ satisfies $D_I D_J \, s = \ft12 \omega_{IJ}\, \square s$, it follows that
$\tilde{\mathcal{Q}}^{(int)}=0$ since $\omega_{IJ} \tilde{C}^{IJ}=0$, and
so one recovers the result \cite{BarTro} that $\mathcal{Q}=
\mathcal{Q}^{(int)}$ for the $\ell=0$ and $\ell=1$ charges
that correspond to the Bondi-Sachs 4-momentum.

In general, however, for an arbitrary function $s$ on the sphere, 
the $ \tilde{\mathcal{Q}}^{(int)}$ are \emph{bona fide} asymptotic 
charges in their own right, which 
supplement the already known BMS charges, $\mathcal{Q}^{(int)}$.
Together, $\mathcal{Q}^{(int)}$ and $-\tilde{\mathcal{Q}}^{(int)}$ provide
the real and imaginary parts of the generalised charges
$\mathcal{Q}$ defined in (\ref{BMS:charge}).

\section{Discussion} \label{sec:dis}

We have shown that one can define new dual asymptotic charges at null 
infinity. These charges are the imaginary part of the charges defined in equation \eqref{BMS:charge}---the real part being the charges of Barnich-Troessaert \cite{BarTro}. The new charges can be defined because at leading order it is 
possible to ``dualise'' the Barnich-Brandt 2-form to obtain an expression 
that also vanishes on-shell. In Ref.~\cite{fakenews}, it was shown that five of the ten conserved non-linear Newman-Penrose charges are subleading charges in the Barnich-Brandt formalism. It is, however, not possible to define dual Barnich-Brandt charges away from null infinity hence the question of how to fit the other five Newman-Penrose charges in
the Barnich-Brandt formalism remains an open problem.

The existence of a further infinite number of BMS charges does not seem to give rise to new soft theorems~\cite{Strominger:2013jfa, He:2014laa} as the imaginary part of $\psi_2^0$ at $\scri^+_{\pm}$ and $\scri^-_{\pm}$ is not part of the physical phase space \cite{He:2014laa}. However, we are nevertheless left with the question of the role of these charges in connection with the information paradox~\cite{Hawking:2016msc,Hawking:2016sgy,Haco:2018ske}.

Dualising the Barnich-Brandt prescription only works for supertranslation charges and at null infinity. In particular, for the SL(2,$\mathbb{C}$) part of the BMS group,   the analysis of appendix \ref{app:charge} does not go through, that is there are terms at order $r^{0}$ that are neither components of the Einstein equation nor total derivative terms; these terms provide an obstruction to a charge being defined. For the same reason, we cannot also understand the imaginary part of the extended BMS charges \cite{fakenews} in this way. It would, therefore, be helpful to understand why it was possible to define dual charges for supertranslations in terms of a more basic Iyer-Wald \cite{IW} (see also Ref.\ \cite{WZ}) or Barnich-Brandt \cite{BB} type of analysis.

\section*{Acknowledgements}

We would like to thank the Mitchell Family Foundation for hospitality at the Brinsop Court workshop where this work was initiated.  
M.G.\ is partially supported by grant no.\ 615203 from the European Research Council under the FP7.  C.N.P.\ is partially supported by DOE grant DE-FG02-13ER42020.

\appendix

\section{Boundary terms} \label{app:charge}
In this section, we prove that the variation of the dual charge \eqref{BB:dual} is equivalent to the Einstein equation.  Starting from the definition of the charge, given in equation \eqref{BB:dual}, and the fact that
\begin{equation}
 \delta g_{ab} = 2 \nabla_{(a} \xi_{b)},
\end{equation}
a calculation similar to that done in appendix D of Ref.\ \cite{fakenews} finds that
\begin{gather} 
  \ndelta \tilde{\mathcal{Q}}= \frac{1}{16 \pi G} \lim_{r \rightarrow \infty} \int_{S}\, d \theta \, d \phi \, \delta^{IJ}_{\theta \phi} \ \Big\{6 \xi_J [\nabla_I, \nabla_c] \xi^{c} +  2  [\nabla_J, \nabla_c] (\xi_I \xi^c) \hspace{35mm} \notag   \\[2mm]
  \hspace{70mm}  + 2 \nabla_c (\xi_I \nabla^c \xi_J + \xi^c \nabla_J \xi_I - \xi_I \nabla_J \xi^c) \Big\}. \label{BB:4}
\end{gather}
Ignoring the first line in the expression above for now, the terms on the second line can be written as
\begin{equation} \label{bdry:1}
 \frac{1}{8 \pi G} \lim_{r \rightarrow \infty} \int_{S}\, d \theta \, d \phi \, \delta^{IJ}_{\theta \phi} \ \nabla_c X_{IJ}{}^c,
\end{equation}
where
\begin{equation} \label{X}
 X_{ab}{}^c =  \xi_a \nabla^c \xi_b + \xi^c \nabla_b \xi_a - \xi_a \nabla_b \xi^c.
\end{equation}
Expanding out the integrand in equation \eqref{bdry:1} gives
\begin{equation}
  \nabla_c X_{IJ}{}^c = \partial_K X_{IJ}{}^K + \partial_{\hat{c}} X_{IJ}{}^{\hat{c}} + \Gamma^c_{ce} X_{IJ}{}^e - \Gamma^e_{dI} X_{[eJ]}{}^d - \Gamma^e_{dJ} X_{[Ie]}{}^d,
\end{equation}
where we use the notation that hatted lower case Latin indices, such as $\hat{c}$, denote $u$ or $r$ components.  The first term is a boundary term and can, therefore, be ignored.  Thus,
\begin{equation} \label{int:1}
 \nabla_c X_{IJ}{}^c = \partial_{\hat{c}}(g^{\hat{c} \hat{d}} X_{IJ \hat{d}}) + \frac{1}{2} g^{-1} g^{\hat{c} d} \partial_d g \, X_{IJ \hat{c}} 
 - 2 g^{\hat{c} d} \Gamma^K_{dI} X_{[KJ]\hat{c}} - 2 g^{c d} \Gamma^{\hat{e}}_{dI} X_{[\hat{e}J]c},
\end{equation}
where
\begin{equation} \label{det:met}
g \equiv \textup{det}(g_{ab}) = - r^4 e^{4\beta} \sin^2 \theta,
\end{equation}
we have used the fact that the equation above is contracted with $\delta^{IJ}_{\theta \phi},$ i.e.\ that the $IJ$ indices are antisymmetrised and also the fact that
\begin{equation}
 X_{[IJ]K} = X_{[IJK]}= 0,
\end{equation}
which can be simply verified from the definition of $X_{abc}$, equation \eqref{X}.  Also, note that
\begin{equation} \label{Xprop:1}
 X_{[ab]c} = \xi_{[a} \partial_{|c|} \xi_{b]} - \xi_c \partial_{[a} \xi_{b]} - \xi_{[a} \partial_{b]} \xi_c = 3 \xi_{[a} \partial_{c} \xi_{b]}
\end{equation}
since the Christoffel symbols cancel out. Moreover, as a direct consequence of the previous equation
\begin{equation} \label{Xprop:2}
 X_{[ab]c} = X_{[ca]b} = X_{[bc]a}.
\end{equation}

 Consider the last term in equation \eqref{int:1}
\begin{equation}
 - 2 g^{c d} \Gamma^{\hat{e}}_{dI} X_{[\hat{e}J]c} = - 2 g^{K d} \Gamma^{\hat{e}}_{dI} X_{[\hat{e}J]K} - 2 g^{\hat{c} d} \Gamma^{\hat{e}}_{dI} X_{[\hat{e}J]\hat{c}}.
\end{equation}
First, we argue that the last term in the expansion above is an order $1/r$ quantity as follows:
\begin{align}
 - 2 g^{\hat{c} d} \Gamma^{\hat{e}}_{dI} X_{[\hat{e}J]\hat{c}} &= - 2 g^{\hat{c} d} \Gamma^{\hat{e}}_{dI} X_{[\hat{c}\hat{e}]J} \notag \\
                                                               &= - 4 g^{d [u} \Gamma^{r]}_{dI} X_{[ur]J},
\end{align}
where in the first equality above we used property \eqref{Xprop:2}.  Using the fact that
\begin{gather} 
\xi_u = -\frac{1}{2} (\Box s + 2 s) + O(1/r), \quad \xi_r = -s + O(1/r^2), \notag\\[2mm] \label{xid}
\xi_I = - r \partial_I s + \frac{1}{2} \left(s D^IC_{IJ} - D^I s C_{IJ} \right) + O(1/r),
\end{gather}
it is clear that
\begin{equation}
 X_{[ur]J} = O(r^0).
\end{equation}
Moreover, using the expression for the Christoffel symbols given in section 4.3 of Ref.\ \cite{BarnichAspects}, 
\begin{align}
 - 4 g^{d [u} \Gamma^{r]}_{dI} &= 2 e^{-4\beta} g_{IL} \partial_{r} C^L \notag \\
                                                               &= O(1/r).
\end{align}
Hence, we find that
\begin{align}
 \delta^{IJ}_{\theta \phi} \nabla_c X_{IJ}{}^c = \partial_{\hat{c}}(g^{\hat{c} \hat{d}} X_{[\theta \phi] \hat{d}}) + \frac{1}{2} g^{-1} g^{\hat{c} d} \partial_d g \, X_{[\theta \phi] \hat{c}} 
 &+ 2 g^{\hat{c} d} \delta^{IJ}_{\theta \phi} \Gamma^K_{dI} X_{[JK]\hat{c}} \notag \\ \label{int:2}
 &- 2 g^{K d}\delta^{IJ}_{\theta \phi} \Gamma^{\hat{c}}_{dI} X_{[JK]\hat{c}} + O(1/r).
\end{align}
Note that
\begin{align}
 \delta^{IJ}_{\theta \phi} X_{[JK]\hat{c}} &= 2 \delta^{IJ}_{\theta \phi} \delta_{JK}^{\theta \phi} X_{[\theta \phi] \hat{c}} \notag \\
                                            &= -\frac{1}{2} \delta^I_K X_{[\theta \phi] \hat{c}}.
\end{align}
Thus,
\begin{equation} \label{int:3}
 \delta^{IJ}_{\theta \phi} \nabla_c X_{IJ}{}^c = \partial_{\hat{c}}(g^{\hat{c} \hat{d}} X_{[\theta \phi] \hat{d}}) + \Big[ \frac{1}{2} g^{-1} g^{\hat{c} d} \partial_d g 
 - g^{\hat{c} d} \Gamma^I_{dI} + g^{I d} \Gamma^{\hat{c}}_{dI}\Big] X_{[\theta \phi]\hat{c}} + O(1/r).
\end{equation}
Now, consider the last two terms in the square brackets above
\begin{align}
  - g^{\hat{c} d} \Gamma^I_{dI} + g^{I d} \Gamma^{\hat{c}}_{dI} &= g^{\hat{c} e} g^{Id} \left( \partial_d g_{Ie} - \partial_e g_{Id} \right) \notag \\
                                                                &= g^{\hat{c} e} g^{Ir} \partial_r g_{Ie} + g^{\hat{c} K} g^{IJ} \left( \partial_J g_{IK} - \partial_K g_{IJ} \right) 
                                                                + g^{\hat{c} \hat{e}} g^{IJ} \left( \partial_J g_{I\hat{e}} - \partial_{\hat{e}} g_{IJ} \right) \notag \\
                                                                &= g^{\hat{c} \hat{e}} g^{IJ} \left( \partial_J g_{I\hat{e}} - \partial_{\hat{e}} g_{IJ} \right) + O(1/r^2)\notag \\
                                                                &= g^{\hat{c} u} g^{IJ} \partial_J g_{Iu} -g^{\hat{c} \hat{e}} \frac{h^{IJ}}{r^2} \partial_{\hat{e}} (r^2 h_{IJ})  + O(1/r^2)\notag \\
                                                                &= - \frac{4}{r} g^{\hat{c} \hat{e}} \partial_{\hat{e}} r  + O(1/r^2), \label{int:4}
\end{align}
where, in the last equality, we have used the fact that $h\equiv \textup{det}(h_{IJ}) = \sin^2 \theta$.  Using equations \eqref{Xprop:1} and \eqref{xid}, one can show that
\begin{align}
 X_{[\theta \phi]u} &=- \frac12 \,  r \,  \delta_{\theta \phi}^{IJ} \Big\{ \partial_{I} s \Big[ s D^K \partial_u C_{JK} - D^K s \partial_u C_{JK} \Big] + \partial_{I} s \partial_{J} \Box s  \Big\} + O(1), \label{X34u}  \\
 X_{[\theta \phi]r} &= \frac{1}{2} \, \delta_{\theta \phi}^{IJ} \,  s \,  \partial_{I} \Big[ s D^K C_{JK} - D^K s C_{JK} \Big] + O(1/r), \label{X34r}
\end{align}
which means the $O(1/r^2)$ terms in equation \eqref{int:4} can be consistently neglected.  Thus,
\begin{equation} \label{int:5}
 \delta^{IJ}_{\theta \phi} \nabla_c X_{IJ}{}^c = \partial_{\hat{c}}(g^{\hat{c} \hat{d}} X_{[\theta \phi] \hat{d}}) + \Big[ \frac{1}{2} g^{-1} g^{\hat{c} d} \partial_d g 
 - \frac{4}{r} g^{\hat{c} r} \Big] X_{[\theta \phi]\hat{c}} + O(1/r).
\end{equation}
Now, 
\begin{equation}
 \frac{1}{2} g^{-1} g^{\hat{c} d} \partial_d g = \frac{1}{2} g^{\hat{c} d} \frac{\partial_d (r^4 e^{4\beta} \sin^2 \theta)}{r^4 e^{4\beta} \sin^2 \theta} = \frac{2}{r} g^{\hat{c} r} + O(1/r^2)
\end{equation}
and
\begin{align}
 \partial_{\hat{c}}(g^{\hat{c} \hat{d}} X_{[\theta \phi] \hat{d}}) &= \partial_{u}(g^{u r} X_{[\theta \phi] r}) + \partial_{r}(g^{ur} X_{[\theta \phi] u}) + O(1/r) \notag \\
                                                                   &= - \partial_{u}( X_{[\theta \phi] r}) - \partial_{r}( X_{[\theta \phi] u}) + O(1/r) \notag \\
                                                                   &= - \partial_{u}( X_{[\theta \phi] r}) - \frac{X_{[\theta \phi] u}}{r} + O(1/r),
\end{align}
so that
\begin{equation} \label{int:6}
 \delta^{IJ}_{\theta \phi} \nabla_c X_{IJ}{}^c = -  \partial_{u}( X_{[\theta \phi] r}) + \frac{X_{[\theta \phi] u}}{r} + O(1/r).
\end{equation}
From equations \eqref{X34u} and \eqref{X34r}
\begin{equation} \label{int:7}
 \delta^{IJ}_{\theta \phi} \nabla_c X_{IJ}{}^c = - \frac{1}{2} \, \delta_{\theta \phi}^{IJ} \, \partial_{I} \Big\{ s \Big[ s D^K  \partial_u C_{JK} - D^K s  \partial_u C_{JK} \Big] + s \partial_{J} \Box s \Big\} + O(1/r).
\end{equation}
In summary, up to total derivative terms, which vanish upon integration
\begin{equation}
  \delta^{IJ}_{\theta \phi} \nabla_c X_{IJ}{}^c = O(1/r).
\end{equation}
Going back to equation \eqref{BB:4} and using the fact that
\begin{equation}
 [\nabla_a,\nabla_b] V_c = R_{abc}{}^d V_d,
\end{equation}
\begin{equation}
 \ndelta \tilde{\mathcal{Q}} = \frac{1}{4 \pi G}  \lim_{r \rightarrow \infty} \int_{S}\, d \theta \, d \phi \, \ \xi_{[\theta} G_{\phi] c} \xi^c,
\end{equation}
where $G_{ab}$ is the Einstein tensor. Hence $\ndelta \tilde{\mathcal{Q}}$ vanishes on-shell at leading order.

\section{Derivation of dual charge} \label{app:dualH}
In this appendix, we show that
\begin{equation}
 \lim_{r \rightarrow \infty} \frac{ H_{\theta \phi}}{\sin \theta} = \frac{1}{2} \Big\{ \delta \big( -  s D_I D_J \tilde{C}^{IJ} \big) + \frac{s}{2} \partial_u C_{IJ} \delta \tilde{C}^{IJ} \Big\},
\end{equation}
where $H_{ab}$ is defined in equation \eqref{H}.  Note that
\begin{equation}
 \frac{ H_{\theta \phi}}{\sin \theta} = \delta^{IJ}_{\theta \phi} \frac{ H_{IJ}}{\sin \theta} = \frac{1}{2} \epsilon^{IJ} H_{IJ}.
\end{equation}
Thus, from equation \eqref{H}, 
\begin{align} \label{Hthph}
\frac{ H_{\theta \phi}}{\sin \theta} = \frac{1}{2} \epsilon^{IJ} \Big\{ \xi_J g^{cd} \nabla_I \delta g_{cd} + \frac{1}{2} g^{cd} \delta g_{cd} \nabla_J \xi_I  -\xi_J \nabla^c \delta g_{Ic} + & \xi^c \nabla_J \delta g_{Ic} \notag \\[3pt]
& + \frac{1}{2} \delta g_{Jc} (\nabla_I \xi^c - \nabla^c \xi_I) \Big\}.
\end{align}

Equations \eqref{BMSgen}, with the fall-off conditions \eqref{met:falloff}, and \eqref{xid} give that
\begin{gather}
 \xi^u = s, \quad \xi^r = \frac{1}{2} \Box s + O(1/r), \quad \xi^I = -\frac{1}{r} D^I s + O(1/r^2), \notag \\ \label{BMSgen:falloff}
 \xi_u = - \frac{1}{2} (\Box s + 2s) + O(1/r), \quad \xi_r =  -s + O(1/r^2), \quad \xi_I = -r D_I s + O(r^0), 
\end{gather}
where $I,J,\ldots$ indices are lowered (raised) with the (inverse) metric on the round 2-sphere.  Consider the first two terms in the expression on the right hand side of equation \eqref{Hthph}.  
Using the expression for the determinant of metric given in equation \eqref{det:met} and assuming implicitly the antisymmetrisation in $[IJ]$, 
\begin{align}
 \xi_J g^{cd} \nabla_I \delta g_{cd} + \frac{1}{2} g^{cd} \delta g_{cd} \nabla_J \xi_I &= \xi_J \partial_I (g^{-1} \delta g) + \frac{1}{2} g^{-1} \delta g \partial_J \xi_I \notag \\
 &= 4 \xi_J \partial_I  \delta \beta + 2 \delta \beta \partial_J \xi_I \notag \\
 &= O(1/r),
\end{align}
where we have used the fall-off properties given in equations \eqref{met:falloff} and \eqref{BMSgen:falloff}.  Hence, these terms will not contribute to the dual charge.
Similarly, using equations \eqref{met:falloff} and \eqref{BMSgen:falloff} and the expression for the Christoffel symbols given in section 4.3 of Ref.\ \cite{BarnichAspects}, it is fairly straighforward to show that
\begin{equation}
 \epsilon^{IJ} \delta g_{Jc} (\nabla_I \xi^c - \nabla^c \xi_I) = O(1/r).
\end{equation}
Hence,
\begin{equation}
\frac{ H_{\theta \phi}}{\sin \theta} = \frac{1}{2} \epsilon^{IJ} \Big\{ \xi_I \nabla^c \delta g_{Jc} +\xi^c \nabla_J \delta g_{Ic} \Big\} + O(1/r).
\end{equation}
Now, consider the first term above:
\begin{align}
 \epsilon^{IJ} \xi_I \nabla^c \delta g_{Jc} &= \epsilon^{IJ} \xi_I \left( g^{ur} \nabla_r \delta g_{uJ} + g^{KL} \nabla_K \delta g_{JL} \right) + O(1/r) \notag \\
                                            &= \epsilon^{IJ} \xi_I \Big[ g^{KL} (\partial_K \delta g_{JL} - 2 \Gamma^M_{K(J} \delta g_{L)M}- 2 \Gamma^u_{K(J} \delta g_{L)u}) -  g^{ur} \Gamma^K_{rJ} \delta g_{uK} \Big]
                                            + O(1/r) \notag \\
                                            &= \epsilon^{IJ} D_I s\ \delta (D^K C_{JK} + 2 C_{0\, J}) + O(1/r).
\end{align}
However, note that equation \eqref{C0eqn} then implies that, in fact,
\begin{equation}
 \epsilon^{IJ} \xi_I \nabla^c \delta g_{Jc} = O(1/r).
\end{equation}
Thus, as before, making use of equations \eqref{met:falloff} and \eqref{BMSgen:falloff} and the expression for the Christoffel symbols given in section 4.3 of Ref.\ \cite{BarnichAspects}
\begin{align}
\frac{ H_{\theta \phi}}{\sin \theta} &= \frac{1}{2} \epsilon^{IJ} \xi^c \nabla_J \delta g_{Ic} + O(1/r) \notag \\
                &= \frac{1}{2} \epsilon^{IJ} \left( \xi^u \nabla_J \delta g_{uI} + \xi^K \nabla_J \delta g_{IK}\right) + O(1/r) \notag \\
                &= \frac{1}{2} \epsilon^{IJ} \left( D_I (s \delta C_{0\, J}) + D^K (s D_I \delta C_{JK}) -  s D_K D_I \delta C_J{}^K + \frac{1}{2} s \partial_u C_{IK} \delta C_J{}^K \right) + O(1/r) \notag \\
                &= \frac{1}{2} \left( -  s D_I D_J \delta \tilde{C}^{IJ} + \frac{s}{2} \partial_u C_{IJ} \delta \tilde{C}^{IJ} \right) + O(1/r),
\end{align}
where in the last equality above we have neglected total derivative terms, which will integrate to zero and have used definition \eqref{Ctwist}.

In conclusion, we find that
\begin{equation}
 \lim_{r \rightarrow \infty} \frac{ H_{\theta \phi}}{\sin \theta} = \frac{1}{2} \Big\{ \delta \big( -  s D_I D_J \tilde{C}^{IJ} \big) + \frac{s}{2} \partial_u C_{IJ} \delta \tilde{C}^{IJ} \Big\}.
\end{equation}

\section{Alternative definition of a real charge} \label{app:C}
 
From the perspective of the Newman-Penrose formalism, it would also make sense to define
charges $\hat Q$ according to
\be
\hat Q= -\fft1{4\pi G}\, \int d\Omega\, s\,
\Big[\psi_2^0 + \sigma^0\, \del_u\, \bar\sigma^0 + \eth^2\, \bar\sigma^0\Big].
\ee
Then, from (\ref{Qdot}), one has
\be
\del_u\, \hat Q = - \frac{1}{4\pi G} \int d\Omega\ s\;
        |\del_u\, \sigma^0|^2 ,
\ee
which holds for any choice of $s$ and demonstrates more clearly that the charges are conserved for vanishing Bondi news. 

 We may re-express $\hat Q$ in terms of the metric expansion coefficients in
(\ref{met:falloff}).  Noting first that $\eth^2\bar\sigma^0= \ft14 D_I D_J(
C^{IJ} + i \, \tilde C^{IJ})$, we find that
\be
\hat Q = -\fft1{8\pi G}\, \int d\Omega\, s\, (F_0 + \ft12
D_I D_J\, C^{IJ}).
\ee
Thus $\hat Q$ is in fact purely real, although it differs from the
real part of $\mathcal{Q}$ defined in (\ref{BMS:charge}), which is given by
$\mathcal{Q}^{(int)}$ in \eqref{BB:Q0} (see equation \eqref{BBrealNP}), by the addition of the second term.

\bibliographystyle{utphys}
\bibliography{NP}

\end{document}